\documentclass[3p,twocolumn]{article}
\usepackage{graphicx}
\graphicspath{ {./figures/} }
\usepackage{hyperref}
\usepackage{float}
\usepackage{verbatim}
\usepackage{color}
\usepackage{subfigure}
\usepackage{amsmath}
\usepackage{amsfonts}
\usepackage{amssymb}
\usepackage{algorithm}
\usepackage{algpseudocode}
\usepackage{algorithmicx}
\usepackage{multirow}
\usepackage[utf8]{inputenc}
\usepackage{booktabs}
\newcommand{\minisection}[1]{\vspace{0.0in} \noindent {\bf #1}\ }

\usepackage{soul}
\usepackage{pifont}
\newcommand{\cmark}{\ding{51}}%
\newcommand{\xmark}{\ding{55}}%

\usepackage{setspace}
\usepackage{soul}
\usepackage{authblk}
\usepackage{threeparttable}
\usepackage{makecell}

\usepackage[hang,flushmargin]{footmisc} 

\renewenvironment{abstract}{%
  \small
  \begin{center}%
    {\bfseries \abstractname\vspace{-0.5em}\vspace{0pt}}%
  \end{center}%
  \quotation}
  {\endquotation}

\begin{document}

\title{Predicting Tweet Engagement with Graph Neural Networks}

\author{Marco Arazzi$^{1,*}$, Marco Cotogni$^{1,*}$, Antonino Nocera$^{1,*}$, and Luca Virgili$^{2,*}$ \\
$^{1}$ Department of Electrical, Computer and Biomedical Engineering, University of Pavia \\
$^{2}$ DII, Polytechnic University of Marche \\
\{marco.arazzi, marco.cotogni\}01@universitadipavia.it\\antonino.nocera@unipv.it, luca.virgili@univpm.it}
\date{}

\twocolumn[\maketitle 
\begin{abstract}
\noindent Social Networks represent one of the most important online sources to share content across a world-scale audience.
In this context, predicting whether a post will have any impact in terms of engagement is of crucial importance to drive the profitable exploitation of these media.
In the literature, several studies address this issue by leveraging direct features of the posts, typically related to the textual content and the user publishing it.
In this paper, we argue that the rise of engagement is also related to another key component, which is the semantic connection among posts published by users in social media.
Hence, we propose {\em TweetGage}, a Graph Neural Network solution to predict the user engagement based on a novel graph-based model that represents the relationships among posts. 
To validate our proposal, we focus on the Twitter platform and perform a thorough experimental campaign providing evidence of its quality.
\end{abstract}
\vspace{\baselineskip}]

\noindent \textbf{\emph{keywords:} Graph Neural Networks, Engagement, Social Network, Twitter, Deep Learning}

\section{Introduction}
\let\thefootnote\relax\footnotetext{
$^*$Equal contribution\\
Code: \url{https://github.com/OcraM17/EngagementGNN}}
\label{sec:Introduction}

Social Networks are, nowadays, an established authority in the context of information sharing.
Although originally they have been designed to favor social interaction among people, in recent years, they have become a powerful tool also for companies to boost their capability to reach customers directly and continuously \cite{baumol2016dynamics}.
This is evident in the context of Industry 4.0, in which Social media marketing has been identified as a critical asset to promote a more effective business involvement of the customers \cite{wan2015industrie}.
Furthermore, the advancement of such a paradigm, referred to as Industry 5.0, includes concepts such as ``human-centric'' and ``mass personalization'' of the production chain, implying an even more prominent need to directly and elastically involve the population, their needs, and opinions in the business strategies \cite{xu2021industry}.

Among the other business sectors, marketing is, for sure, one of the drivers of the interactions with the final customers of every company.
Understanding and being able to predict the capability of content published on a social platform to reach the desired audience is, for sure, a timely and well-studied subject for marketing applications \cite{bhattacharya2019viral,cheung2020influence}.

\begin{figure*}
    \includegraphics[width=\textwidth, height=5.5cm]{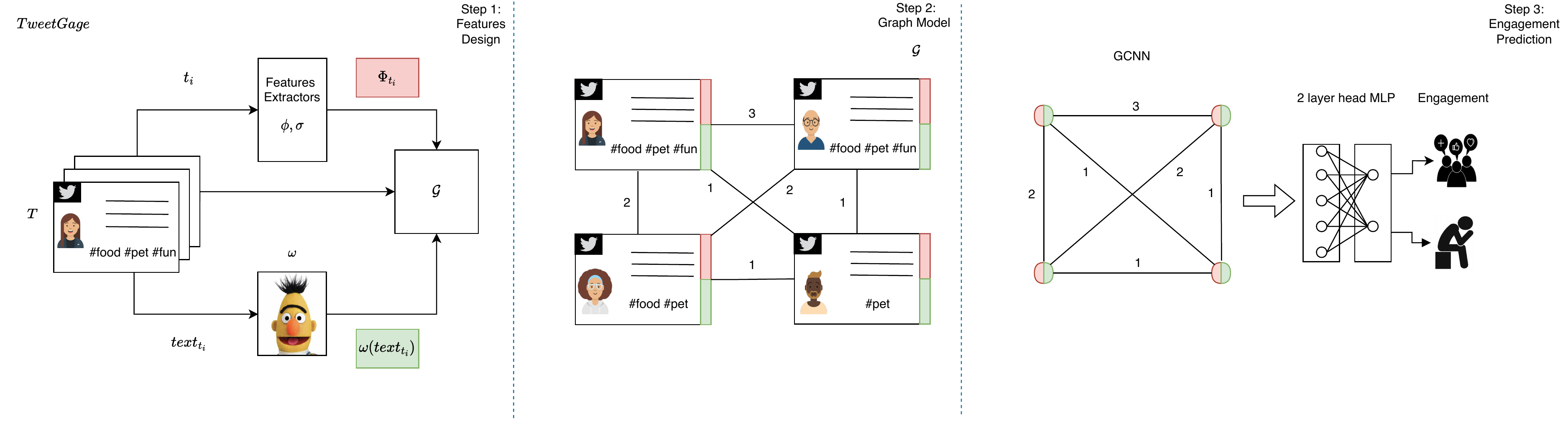}
    \caption{Workflow of the three steps composing {\em TweetGage}. In the first step, given a tweet $t_i \in T$, features associated with the post and the user are retrieved using two feature extractors $\phi$ and $\sigma$. In the same way, given the text of the tweet $text_{t_i}$, an embedding of the text is obtained using a pre-trained language model $\omega$. During the second step, the graph model $\mathcal{G}$ is created connecting posts (nodes) that share at least one hashtag. The features and embedding previously computed are associated with the corresponding node. Finally, in the last step, the model $\mathcal{G}$ is provided as input to a Graph Convolutional Neural Network with the aim of predicting the user engagement of each tweet.}
    \label{fig:teaser}
\end{figure*}

However, if, on the one hand, the capability of content to be spread across the population is of fundamental importance, also the interest generated in the recipients is a crucial aspect to investigate, on the other hand.

Following this intuition, in the recent literature, several studies have been devoted to the definition and analysis of the concept of user engagement in social networks \cite{Dolan*19,aldous2019view}.

Typically, in the context of social media, marketers considered the number of reactions received on a post as a quantitative measure of user engagement \cite{bakhshi2014faces,paine2011measure,suh2010want}.
Of course, how the engagement is computed from a technical point of view, strictly depends on the functionalities provided by the target social medium. However, classical engagement definitions consider the percentage of users who reacted to a post by expressing an opinion or comment, or by directly exhibiting an appreciation (e.g., ``like'', ``retweet'', ``favorite'', and so forth).
In any case, according to the scientific literature, the features that have an impact on the rise of user engagement can be categorized into three main classes: the ones related to the user creating the post, those related to the post context (such as time, group, topic, etc.), and those related to the content itself (text features, presence of media, etc.) \cite{jaakonmaki2017impact}.

From these premises, many researchers have developed solutions, typically based on machine learning, to study the dynamics of user engagement and to build solutions to support companies by predicting whether a content will be capable of generating reactions \cite{Toraman*22, ChPi12, Magalhaes*14}.

However, while these approaches achieved quite satisfactory results, the high dynamics of social networks and the crucial roles of communities in the diffusion of contents guided researchers to identify features more and more related to the structural properties of social media \cite{Purohit*21, Daniluk*21}.

Motivated by this research direction, in this paper we propose {\em TweetGage}, a novel approach for the binary classification of tweets according to user engagement, i.e., for predicting whether published content can generate engagement (a reaction rate non equal to zero) or is destined to remain unnoticed. 
In our study, we argue that, although the network of connections among users and the textual features of the content plays a crucial role, engagement is also triggered by the semantic connections among posts. 
Indeed, intuitively, if a post relates to a sequence of previous posts that are attracting the attention of users, then it will have a higher probability to generate engagement.
To capture this dynamic, in this paper we introduce a graph-based model in which the nodes are the posts and the connection among them encodes the information of whether their content is overlapping to some extent.
To evaluate such overlapping, we exploit the hashtags written by the authors to give weights to the links among posts. 
Then, we leverage the Graph Neural Network (GNN) technology to process such data, along with basic information about the posts as devised by the scientific literature and identify important features that can be used to model the dynamics of engagement generation. The workflow of our solution is visible in Figure \ref{fig:teaser}.

To validate our proposal, we focus on the Twitter platform and perform a deep experimental campaign to assess the quality of our results by comparing them with those achieved by existing approaches.
The motivations underlying the choice of Twitter as referring platform, are to be found in its extremely high popularity as a social media and in the interest that the research community has devoted to it in recent years \cite{malik2019use,xue2020twitter}.
Interestingly, our GNN-based approach outperforms state-of-the-art solutions in all the performed tests.
Finally, our ablation study on the role of the considered features provides useful insights into the dynamics leading to the generation of user engagement in social platforms and demonstrates the correctness of our intuition.

The plan of this paper is as follows. In Section \ref{sec:Related-Work}, we report a review of the related literature. In Section \ref{sec:Methodology}, we describe our proposal and the adopted methodology. In Section \ref{sec:Experiments}, we report the experiments to validate our approach along with our ablation study to assess the role of the considered feature. Finally, in Section \ref{sec:Conclusion} we draw our conclusions and look at possible future works.

\section{Related Work}
\label{sec:Related-Work}

The concept of user engagement in social networks has been heavily studied in the literature \cite{Dolan*19, Shahbaznezhad*21, Read*19}. Indeed, due to the enormous popularity of social media, such as Facebook, Twitter, and TikTok, many companies are increasingly investing in content creation and distribution on these platforms. Evaluating the outcome of these investments is not straightforward, but it is fundamental to understand how to effectively create posts on social platforms. One of the most used metrics to measure the impact of a post is user engagement \cite{Shahbaznezhad*21}.

Depending on the social network, there are many ways to compute the engagement of a post, since users have different actions for interacting with it \cite{Dolan*19}. For instance, on Twitter, the user can like, retweet, and reply to a post, and these actions are considered for developing a user engagement metric \cite{Read*19, Munoz*17, McShane*21, Diaz*14}. In our case, we decided to use the formula of \cite{Diaz*14} for computing the engagement, as we will see in Section \ref{sec:Methodology}.

Due to the high interest of both researchers and companies to study user engagement in social networks, it is not unexpected that many approaches to predict this engagement has been developed over time \cite{ChPi12, Anelli*20, Anelli*21, Mapredicting21}. Many of these approaches leverage Machine Learning and Deep Learning algorithms. From our perspective of using network features, we can identify two lines of thought in the literature: {\em (i)} prediction of engagement without considering graph-based features \cite{Silva*20, Toraman*22, ChPi12, Magalhaes*14, Orellana*16, AlAnJa22}, and {\em (ii)} prediction of engagement with graph-based features \cite{Diaz*14, Volkovs*20, Purohit*21, Daniluk*21, Felicioni*20, Abdollahi*14}.

In the former case, researchers employed features from the user profile, from his/her tweets' performance in terms of the number of likes, retweets, replies, and mentions, and from the analysis of the post text. For instance, in \cite{Toraman*22} the authors analyze sets of features that reflect user behaviors, tweet statistics, and the semantics of text through BERT. They tested their prediction performance with a Light Gradient Boosting Machine (LightGBM) and a Multilayer Perceptron (MLP) in a supervised task. The results highlight that users would engage with tweets based on text semantics and contents regardless of the tweet author, even if the popularity of the user could be useful for replies and mentions. In \cite{Silva*20}, the authors investigated the relationship between misinformation and user engagement in COVID-19-related tweets. They measured the engagement of a post as the sum of likes and retweets and then labeled them as high or low engagement based on the median value of engagement distribution. From the tweet text, they extracted features such as the number of words, Part-Of-Speech tagging, etc. while no deep learning-based embedding was used. The resulting dataset is the input to Gradient Boosting, Multinomial Naive Bayes, and Random Forest classifiers, which achieved very good performances. However, we point out that these kinds of approaches do not consider the interactions among users and/or posts, which represent the core of social networks and from which we can extract information to improve the classification results.

In the latter case, the researchers used both the features from the former case and graph-based features extracted from the users and/or the corresponding tweets. In \cite{Diaz*14}, the authors predict which posts will generate the highest user engagement on Twitter. They modeled the scenario as a Collaborative Ranking task and proposed to use the user-item-tweet interactions. Here, the features were extracted from the tweets and users' statistics, such as the number of tweets posted, average user engagement from their history, the ratio of the number of user friends to the number of his/her followers, etc. They learned a scoring function that directly optimizes user engagement in terms of normalized discounted cumulative gain on the predicted ranking. In \cite{Daniluk*21}, the authors fine-tuned a DistilBERT model \cite{Sahn*19} on tweets to obtain a text embedding and used Efficient Manifold Density Estimator \cite{Dabrowski*21} to represent the same text as a compressed and fixed-size representation of the tweet meaning. Furthermore, they added some features concerning community detection on directed graphs of engaged-engaging user interactions. In this way, they captured potentially complex communities of mutual interaction between users, with large communities having lower interaction strength. These features are fed into a simple shallow feed-forward neural network that predicts tweet engagement. In \cite{Purohit*21}, the authors proposed a framework called People-Content-Network Analysis (PCNA), which is designed to analyze human dynamics on social networks. This framework uses three categories of features: {\em (i)} community features, which are measurements of the community like its size, the total number of active users that another one is following, etc.; {\em (ii)} author features, such as counts of followers, and following; and {\em (iii)} content features, such as a number of retweets, mentions, hashtags, and keywords. Finally, a Support Vector Machine classifier is trained to predict tweet engagement, and they demonstrated that all these categories of features are useful for obtaining good performances. In \cite{Felicioni*20}, the authors proposed an ensemble model composed of two stages: the first one is made up of three LightGBM, Gradient Boosting, and neural networks, while the second one is another LightGBM. The input to this last model consists of two groups of features. The first one concerns the modeling of user behavior, such as the number of active and passive engagements, the number of engagements with language or hashtag, and user similarity. This last is extracted through an undirected graph, where each edge connects two users if one engaged with a tweet created by the other and has a weight equal to the number of such engagements. The second one regards tweet text features, which contain the text embeddings produced by DistilBERT, the unique word frequency (how much a user tends to use the same words over time), and the tweet topic (the authors identified some popular topics and manually associated each to a list of the most used words). 

Even if some authors considered graph-based features to predict the engagement of a post, there are some differences between our approach and other ones. As we will see in Section \ref{sub:Model}, we model Twitter through a network of posts based on hashtags that also have a time threshold to control the number of connections. Then, this network of posts is the input to a Graph Neural Network that will predict the engagement of a tweet. The Machine Learning and Deep Learning algorithms so far employed in the literature are not specifically designed to deal with graphs, which is a limitation that we want to point out. We will show that Graph Neural Networks are suitable for this scenario because they can learn meaningful patterns from the underlying graph of the posts, and then leverage this knowledge to achieve very high performances.

\section{Methodology}
\label{sec:Methodology}

In this section, we describe our methodology to deal with engagement prediction on Twitter. In Section \ref{sub:Model}, we introduce our model to represent the scenario, while in Section \ref{sub:GCNN}, we report a brief description of Graph Convolutional Neural Networks and their importance in our case.

\subsection{Model Proposal}
\label{sub:Model}
In this section, we introduce a suitable network model to represent Twitter posts and their interactions. Let $T$ be a set of Twitter posts, where $t_i \in T$ is a tweet, and $\phi(t_i)$ a function that extracts information about $t_i$, such as timestamp, text, favorite count, the user identifier that wrote it and so on. In our case, $\phi(t_i)$ associates the following features with each $t_i$: 

\begin{itemize}
    \item $id_{t_i}$: the identifier of $t_i$;
    \item $\tau_{t_i}$: the posting timestamp of $t_i$;
    \item $text_{t_i}$: the text corresponding to the tweet $t_i$;
    \item $Length \ of \ post$: the number of characters present in $t_i$;
    \item $Emojis$: the number of emojis used;
    \item $u_{t_i}$: the username of the author of the tweet;
    \item $Has \ media$: if $t_i$ contains an image and/or video;
    \item $Favorite_{t_i}$: the number of likes received by $t_i$;
    \item $Retweet_{t_i}$: the number of times $t_i$ was retweeted;
    \item $Official \ Source$: if $t_i$ was published through Twitter websites or Twitter API instead of third parties;
    \item $h_{t_i}$: the sets of hashtags used in $t_i$;
    \item $Number \ of \ hashtags$: the number of hashtags present in $h_{t_i}$;
    \item $Number \ of \ Mentions$: the number of times a username is mentioned in $text_{t_i}$.
\end{itemize}

In order to process $text_{t_i}$, we need a function that maps the text into a vectorial representation. To this end, we define a function $\omega(text_{t_i})$ as the embedding of $text_{t_i}$, which takes in input a natural text and returns a numeric vector representing such text in a continuous space. The function $\omega(text_{t_i})$ could be a suitable embedding model available in the Natural Language Processing literature, such as Word2Vec, GloVe, or BERT \cite{Otter*20}. 

Moreover, we need some information about the user posting the tweet $t_i$. Recall that $u_{t_i}$ is the user identifier of the author of at least a tweet $t_i \in T$. We define a function $\sigma(u_{t_i})$ that extracts information about a user $u_{t_i}$, such as the number of followers and following, the number of tweets, a boolean value specifying if he/she is a verified user, and so on. Starting from $u_{t_i}$, we leverage $\sigma(u_{t_i})$ in order to extract the following information:
\begin{itemize}
    \item $Verified \ user$: a boolean value representing if $u_{t_i}$ is a verified account or not;
    \item $Followers$: the number of followers of $u_{t_i}$;
    \item $Following$: the number of following of $u_{t_i}$;
    \item $Number \ of \ Tweets$: the number of tweets posted by $u_{t_i}$ since the creation of the account.
\end{itemize}

Now, we can define a graph $\mathcal{G}$ to model the tweets contained in $T$ along with their corresponding interactions. Specifically, let $\mathcal{G} = \langle T, E_\delta \rangle$ be a network of tweets. Here, the set of edges is represented by $ E_\delta$, where there is an edge $e_{ij} = (t_i,t_j, w_{ij})$ if the tweet $t_i$ and $t_j$ share at least one common hashtag and if they were published within a specific time interval $\delta$. The weight $w_{ij}$ is equal to the number of common hashtags between $t_i$ and $t_j$. Formally speaking, the set of edges is defined as $E_\delta = \{\langle t_i, t_j, |h_{t_i} \cap h_{t_j}| \rangle \ s.t.\ t_i, t_j \in T, h_{t_i} \cap h_{t_j} \neq \emptyset, |\tau_{t_i} - \tau_{t_j}| < \delta \}$.

In our model, $\delta$ represents a threshold to control the number of edges between the nodes in $\mathcal{G}$. Indeed, on Twitter and other social networks, hashtags are a way to collect posts on similar topics, which sometimes are general (such as \#politic, \#healthcare, etc.) and sometimes are specific (such as \#WorldCup2022, \#Election2022). Defining the duration of a hashtag on a social network is not straightforward, since it could last either a day or months. Tweets that are posted today can contain the same hashtags as older tweets, even if the former regards a newer event in that topic and should not be directly connected to the latter. Following this reasoning, we decided to add a temporal threshold $\delta$ in order to create edges between posts only if they are published within a specific time interval. We can assume that $\delta$ narrows down the topics discussed in tweets and connects them in a smarter way. In the literature, many papers have studied the lifespan of tweets to predict popularity and generate engagement \cite{Linz*21, Zhao*15, Bae*14}. According to the literature, the lifespan of a tweet is between 10 and 30 minutes, and for this reason, we decided to set $\delta=15$ minutes. This means that two posts $t_i$ and $t_j$ are connected by an edge in $\mathcal{G}$ only if they share at least one hashtag and if they were published 15 minutes apart.

Now, we need to define a way for evaluating the engagement of a tweet. To do so, we adopt a formula proposed in \cite{Diaz*14} to compute the engagement of $t_i$:

\begin{equation}
\label{eq:engagement}
eng(t_i) = Favorite_{t_i} + Retweet_{t_i}
\end{equation}

\noindent
where $Favorite_{t_i}$ is the number of favorites received by $t_i$, and $Retweet_{t_i}$ is the number of times $t_i$ was retweeted.

In order to perform binary classification on tweets that did or did not generate engagement, we must add a set of class labels $L$ for each $t_i \in T$. To this end, we define $l_{t_i}$ the label of a post $t_i$ as:

\[   
l_{t_i} = 
     \begin{cases}
       \text{0} &\quad\text{if $eng(t_i) = 0$}\\
       \text{1} &\quad\text{if $eng(t_i) \ge 1$} \\
     \end{cases}
\]

Roughly speaking, a label $l_{t_i}$ equal to $0$ is assigned to all the posts that generated no engagement in terms of favorites and retweets during their presence on Twitter (not only during the $\delta$ threshold). On the other hand, a label $l_{t_i}$ equal to $1$ is associated with all the posts that generated at least one favorite or retweet action by a user.

\subsection{Graph Convolutional Neural Network}
\label{sub:GCNN}
Deep learning models effectively learn hidden patterns of Euclidean data, but there are many scenarios in which data are represented through graphs \cite{Wu*20}. Networks are hard to process since they could have a variable number of nodes, which, in turn, could have a different number of neighbors. This means that typical operations of deep learning, like convolutions, are difficult to apply to the graph domain. Moreover, instances in a network (i.e., nodes) are related to each other by links of various natures, which violates the assumption of machine learning of independent instances in the dataset. All these issues are addressed by Graph Neural Networks (GNN) \cite{Scarselli*08}, which extracts high-level representations of nodes and edges of a graph given as input. In \cite{Bruna*13} the authors introduced convolution operations on graphs thanks to the spatial construction algorithm and the spectrum of the graph-Laplacian. Convolutions on graphs allow the learning of node embeddings that consider also neighborhood information, which provides much richer representations. GNNs are used in many different tasks, such as node classification, graph classification, link prediction, and graph clustering. 

Since we modeled the Twitter scenario through a graph $\mathcal{G}$ representing the tweets and their interactions, and we wanted to evaluate the engagement of posts according to the network, we decided to use GNN models to perform node classification. To the best of our knowledge, our paper is the first attempt to employ this kind of model to predict the engagement of social posts. From the different GNNs architectures available in the literature \cite{Wu*20}, the one that is suitable for node classification and that can easily work with our dataset is the Graph Convolutional Neural Network (GCNN). The main idea of GCNNs is to apply the convolution operation to graph data, and so to generate a node embedding by aggregating its features and neighbors’ features. Two variants of GCNNs gained a lot of attention lately: GraphSAGE \cite{HaYiLe17} and Graph Attention Network (GAT) \cite{Velickovic*17}. Both these architectures are spatial-based, and the most important difference lies in the assignment of weights to the neighbors of a node. Indeed, GraphSAGE explicitly assigns a non-parametric weight according to the degrees of the nodes involved in the aggregation process, while GAT implicitly computes this weight through the attention mechanism and a neural network model, in order to give larger weights to the most important nodes. As will be clear in the next sections, in our context, GraphSAGE performs better than GAT due to the high peculiarity of the considered scenario, therefore we adopted GraphSAGE as referring GCNN architecture.

According to GraphSAGE \cite{HaYiLe17}, we define $T_{t_i}$ as the set of nodes that has an edge with the node $t_i$, and so it represents the immediate neighborhood of $t_i$. The process to obtain a node embedding is iterative and consists of $n$ steps. At each step $k$, let $r_{t_i}^k$ be the node representation of $t_i$ at this step. We aggregate the representations of the nodes in its neighborhood $r_{T_{t_i}}^k = \{r_{t_j}^{k}, \forall {t_j} \in T_{t_i}\}$. In our case, we use a sum operator to aggregate the representations of the neighbors of $t_i$. Then, as in GraphSAGE \cite{HaYiLe17}, we concatenate the $r_{t_i}^{k}$ with the aggregated neighborhood vector $r_{T_{t_i}}^{k}$, and feed it to a fully connected layer with a GELU activation function. 

According to the representation reported in Section \ref{sub:Model}, the features associated with a node of our GCNN are derived both from the tweet and the user posting it. As for the creator of a tweet, we used {\em Followers}, {\em Number of Tweets}, {\em Following}, while as for the tweet we leveraged {\em Length of a post}, {\em Number of hashtags}, {\em Number of mentions}, {\em Emojis}, {\em Official Source}, {\em Has media}. We define this last set of features as $\Phi_{t_i}$, which includes user and tweet data. Then, we define $\omega(text_{t_i})$ as the embedding of $text_{t_i}$ computed through a pre-trained BERT.


\section{Experiments}
\label{sec:Experiments}

In this section, we present the experiments we carried out to evaluate the proposed model in the engagement prediction task. We begin by describing the dataset for our experiments with the associated statistics in Section \ref{sub:Dataset}.
Following that, in Section \ref{sub:GraphAnalysis}, we present the network analysis of the graph which motivated our strategy of exploiting its intrinsic patterns to improve the engagement prediction of a post. 
Then in Section \ref{sub:ComparisonResults}, we compare our solution with other models that take as input only the information regarding the features extracted from the text of the post and the user who posted it. 
Finally, in Section \ref{sub:Ablation}, we perform an ablation study on the different components of our model to evaluate whether the results obtained are actually due to the advantage of considering the intrinsic information of the graph.

\subsection{Dataset Description}
\label{sub:Dataset}

Our Twitter dataset consisted of the tweets posted during November 2021\footnote{https://archive.org/details/archiveteam-twitter-stream-2021-11}, and was built through a real-time stream using the Twitter API. Once we downloaded the Twitter dataset of November 2021, we performed some data cleaning operations. First of all, we removed all the tweets that were not written in English. Then, since the tweets in the original dataset did not contain any information about the interactions they received (such as the number of likes, mentions, and retweets), we developed a Python script to update the tweets data, using, again, the Twitter APIs \cite{kearney2019rtweet, model2014}. Moreover, due to the very high number of tweets present in the dataset (more than 1 million), we decided to select one week of data, from November $1^{st}$ 2021 to November $7^{th}$ which still remains a large dataset. Finally, we computed the engagement of each tweet according to Equation \ref{eq:engagement}. The statistics of our dataset after the data cleaning operations are reported in Table \ref{tab:dataset-description}.

\begin{table}[h]
    \caption{Statistics of our dataset}
    \vspace{+0.4cm}
    \label{tab:dataset-description}
    \begin{tabular}{lc}
        \toprule
        Number of days & 7 \\
        Number of users & 194,046 \\
        Number of posts & 243,750 \\
        Mean number of posts by day & 34,821.43 \\
        Number of unique hashtags & 94,646\\
        Median number of posts by user & 1 \\
        Max number posts by user & 126 \\
        \bottomrule 
    \end{tabular}
\end{table}

From the analysis of Table \ref{tab:dataset-description}, we observed that our dataset contained $243,750$ tweets spread in the first week of November. There are many different hashtags used in the tweets, which will be useful when we will create the corresponding post network. As in many social network scenarios \cite{IS21, phis12}, we saw that the median number of posts by a user is low (i.e., $1$) while the maximum number of posts is quite high for a week (i.e., $126$). This highlighted that there were many differences in terms of posting activities among authors, in which few users wrote a lot of tweets, while most of them wrote only one or two posts. Starting from our dataset of tweets, we created the corresponding graph ${\mathcal G}$, whose characteristics are reported in Table \ref{tab:graph-statistics}.

\begin{table}[ht]
    \caption{Graph statistics}
    \vspace{+0.4cm}
    \label{tab:graph-statistics}
    \begin{tabular}{lc}
        \toprule
        $\delta$ (minutes) & 15 \\
        Number of nodes & $243,750$\\
        Number of edges & $4,403,434$ \\
        Density & $1.48$ $e^{-4}$ \\
        Number of connected components & $120,434$\\
        Maximum connected component & $18,322$ \\
        \bottomrule 
    \end{tabular}
\end{table}

\begin{figure}[h]
  \centering
  \includegraphics[width=0.45\textwidth]{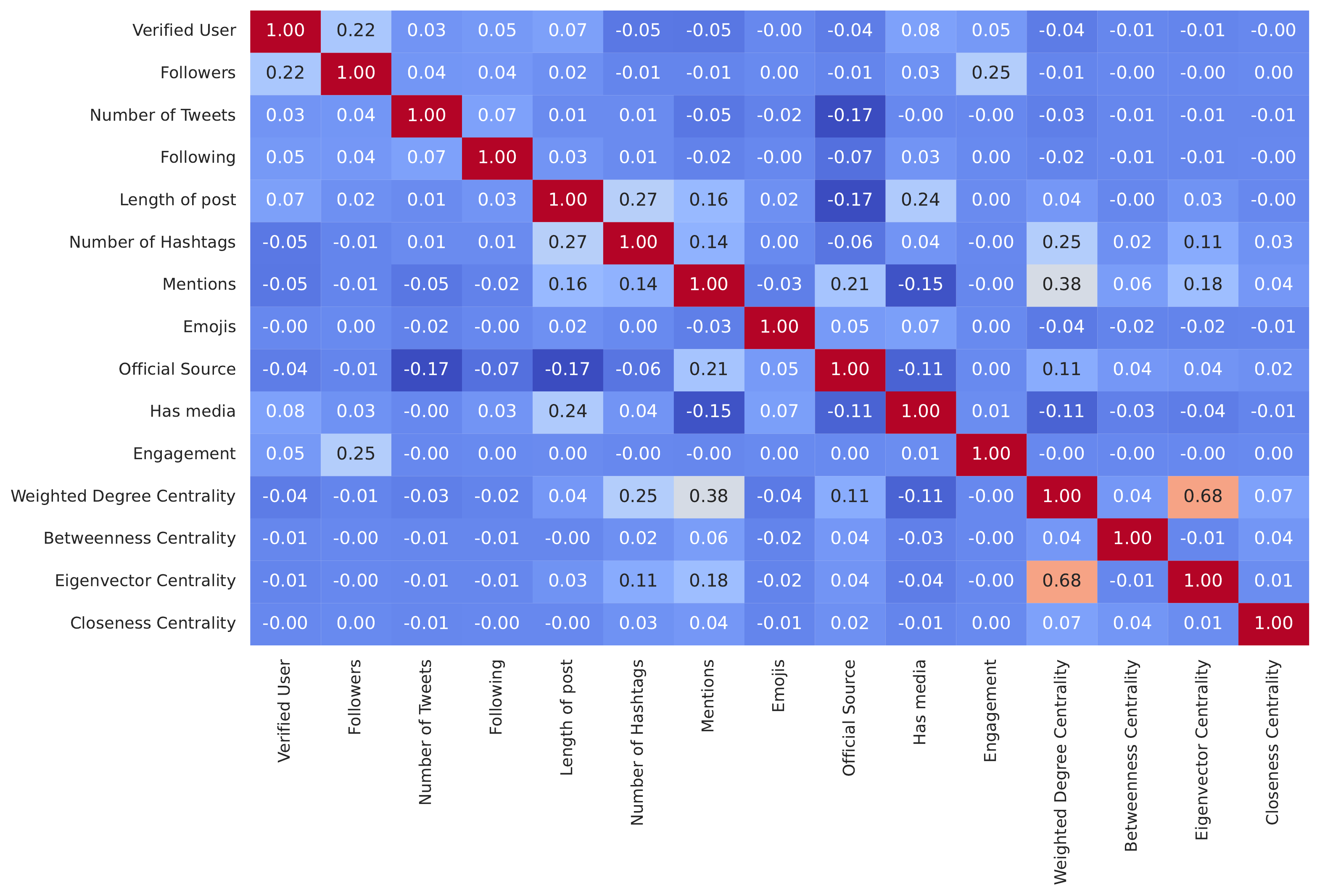}
  \caption{Correlation matrix of the dataset and graph features}
  \label{fig:Correlation-Matrix}
\end{figure}

From Table \ref{tab:graph-statistics}, we observed that the number of nodes corresponds to the number of posts and that we had millions of edges connecting tweets together. It is worth noting that, even if we set a  threshold ($\delta=15$ minutes) to drive the creation of edges, we still obtained a high number of connections. Finally, we saw that there are many connected components ($49.40$\% of the number of nodes), and the maximum component consisted of $18,322$ nodes. This means that there were several isolated nodes and/or components with few nodes, which is a phenomenon both depending on the used hashtags and on our $\delta$ threshold. Once we built ${\mathcal G}$, we computed four centrality measures (i.e., Weighted degree centrality, Closeness centrality, Betweenness centrality, and Eigenvector centrality \cite{buccafurri2013measuring,landherr2010critical}) for each post, and then created a correlation matrix of all the features obtained so far. The results are reported in Figure \ref{fig:Correlation-Matrix}.

From Figure \ref{fig:Correlation-Matrix}, we observed that there were very few interesting correlations between two features in our dataset, and some of them are expected. Specifically, the engagement of a post was correlated with the number of followers of the user. Then, the weighted degree centrality and eigenvector centrality were correlated with the number of hashtags and the mentions present in a post, which can be explained by the construction of ${\mathcal G}$. In conclusion, there are no unexpected correlations in our dataset, and the engagement of a post had poorly correlated with other features. This suggested that there were no linear relationships in the tweet dataset, and so we had to exploit non-linear approaches to predict the engagement of a post.

\subsection{Analysis of Centrality Measures and Engagement}
\label{sub:GraphAnalysis}

\begin{figure}[h]
  \centering
  \includegraphics[width=0.4\textwidth]{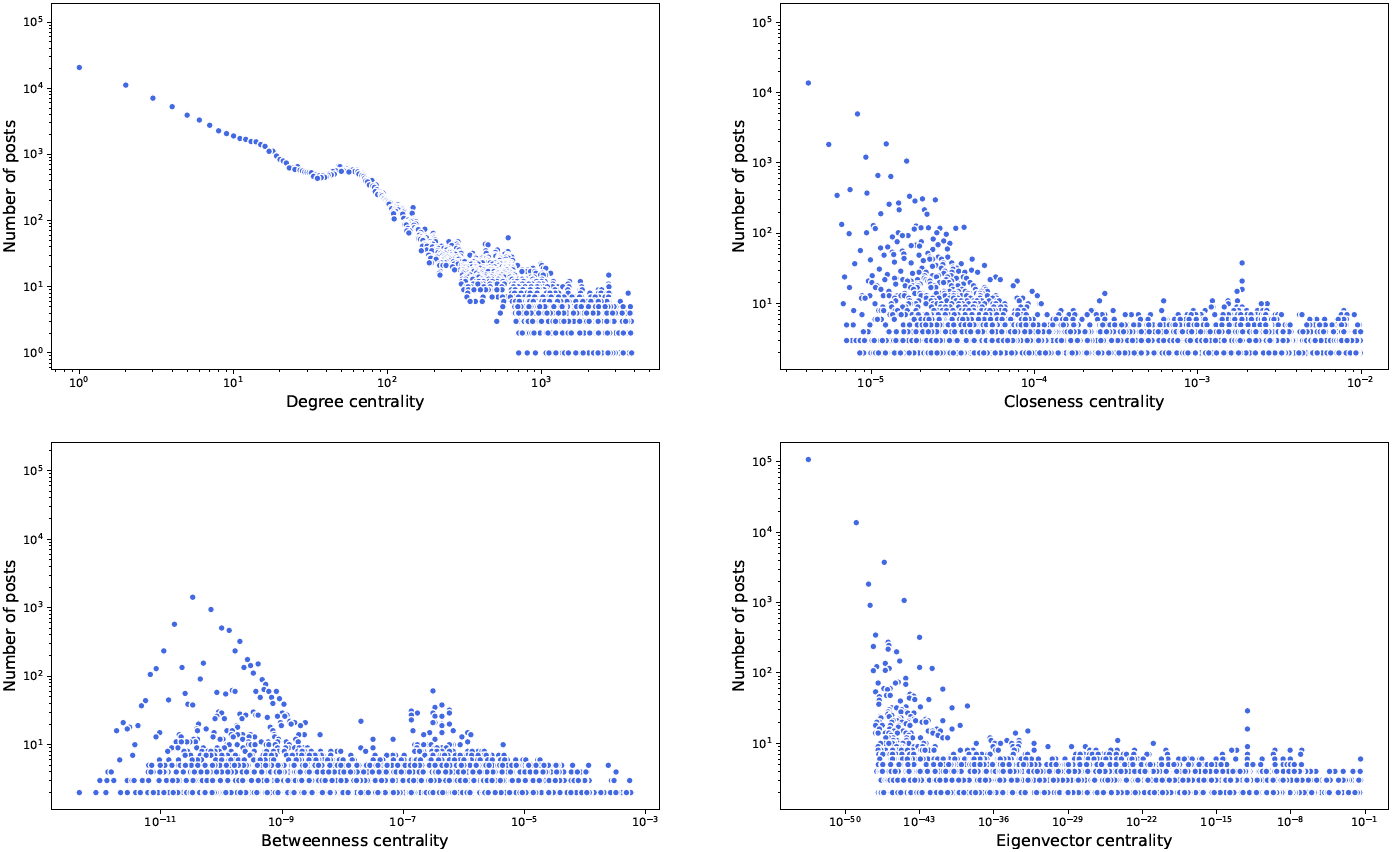}
  \caption{Log-log distribution of the number of posts against the common centralities}
  \label{fig:Centralities}
\end{figure}

We analyzed the network centralities of the posts in order to verify possible differences between posts with engagement and posts without engagement. Recall that, each centrality depicts a different way of describing the role of nodes, such as the number of connections, their impact in the information flow, or their connection with important peers. We considered the four most important centrality measures and plotted the distribution of the obtained results in Figure \ref{fig:Centralities}.

From Figure \ref{fig:Centralities}, we observed that the Weighted Degree, Closeness, and Eigenvector centralities followed a power law distribution, even if the first one was far steeper than the others. On the contrary, the Betweenness centrality had a more equal distribution of values than the previous ones. These distributions highlighted that the posting network reflected the same phenomena of classical social network analysis, so to deepen our analysis we proceeded by studying the differences between the posts with and without engagement. To this end, we split the posts by the class $l_{t_i}$ and computed the distribution of their corresponding centrality values. We reported the summarized results in Table \ref{tab:ks-centralities}.

\begin{table}[h]
    \caption{Centrality measures statistics of the nodes of $\mathcal{G}$ split by engagement classes and results of Kolmogorov–Smirnov tests between the distributions of centrality values of the two classes}
    \vspace{+0.4cm}
    \label{tab:ks-centralities}
    \resizebox{1\linewidth}{!}{\begin{tabular}{lcccccc}
        \toprule
        & $l_p$ & Mean & Std & K-S test statistic & K-S test p-value\\
        \midrule
        \multirow{2}{*}{Weighted Degree Centrality} & 0 & 110.63 & 384.05 & \multirow{2}{*}{0.15} & \multirow{2}{*}{< 0.01 }\\
        & 1 & 39.12 & 197.26 & & \\
        \multirow{2}{*}{Closeness Centrality} & 0 & 1.12 $e^{-3}$ & 2.14 $e^{-3}$ & \multirow{2}{*}{0.02} & \multirow{2}{*}{ < 0.01 }\\
        & 1 & 1.06 $e^{-3}$ & 2.09 $e^{-3}$ & & \\
        \multirow{2}{*}{Betweenness Centrality} & 0 & 3.51 $e^{-6}$ & 2.13 $e^{-5}$ & \multirow{2}{*}{0.14} & \multirow{2}{*}{ < 0.01 } \\
        & 1 & 2.28 $e^{-6}$ & 1.91 $e^{-5}$ & & \\
        \multirow{2}{*}{Eigenvector Centrality} & 0 & 2.38 $e^{-4}$ & 1.08 $e^{-3}$ & \multirow{2}{*}{0.15} & \multirow{2}{*}{ < 0.01} \\ 
        & 1 & 4.53 $e^{-5}$ & 1.10 $e^{-3}$ & & \\
        \bottomrule 
   \end{tabular}}
\end{table}

As for the Weighted Degree Centrality, nodes with $l_{t_i} = 0$ had a higher mean and standard deviation than nodes with $l_{t_i} = 1$. This means that posts without engagement were much more connected than posts with engagement, which also leads to the conclusion that having a higher number of hashtags in a tweet is not a way to create engagement. As for Closeness and Betweenness centralities, there was no difference as in the previous case. In the Eigenvector centrality case, the mean of posts with $l_{t_i} = 0$ was higher by one order of magnitude w.r.t. the mean of posts with $l_{t_i} = 1$. In order to statistically verify the discrepancies between these distributions, we ran a Kolmogorov-Smirnov test and reported the results of the test statistics and p-values in the same table. Since the p-values were always lower than 0.05, we stated that the distributions of network centralities of posts with $l_{t_i} = 0$ and $l_{t_i} = 1$ had statistically different, and so that posts with engagement have specific patterns w.r.t. posts without engagement. This finding supported our intuition that a machine-learning approach capable of learning on graphs is useful to predict the engagement of a post in this particular scenario.


\subsection{Results}
\label{sub:ComparisonResults} 

The advantage of our proposal over more traditional ones is the fact that it exploits the intrinsic patterns of the graph in addition to the classical features directly derived from the posts. To assess the quality of our proposal, we considered a set of baseline architectures that take as input only the features related to the single posts. 
In the baselines, we included the solution that won the $RecSys \ Challenge$ in 2020 \cite{Anelli*20} whose goal was the same as the one that we are addressing in this paper (XGBoost experiment, in the following).
In addition, since in our approach, we combine the features of the post with a vectorial representation of the corpus obtained by a pre-trained BERT-based model, we decided to perform a fine-tuning of such BERT model for our specific task, to verify if the features of the text alone are enough to predict the engagement (BERT FT experiment).
Then, we took into consideration two solutions based on neural networks: a Multilayer Perceptron (MLP experiment) and a Convolutional Neural Network (CNN experiment), characterized by a one-dimensional Convolution Layer followed by a fully connected classifier.



Finally, as stated in Section \ref{sub:GCNN}, we included in the comparison a variation of our proposal using Graph Attention Networks (GAT experiment).
In detail, this variation of our approach used a Multi-Head Attention Layer instead of the Convolutional ones. In Table \ref{tab:results}, we report the results of the experiments.

\minisection{Implemention details.}For our experiments, we used TensorFlow \cite{abadi2016tensorflow} as Python framework. The BERT FT experiment was performed by training a linear layer using the features extracted from the posts (768) as input. For the MLP experiment, we build a two layer MLP network with $32$ hidden neurons. As for the CNN experiment, we considered a convolutional neural network composed of two 1D convolutional layers. For the XGBoost experiment, instead, we considered the solution that won the $RecSys \ Challenge$ in 2020 \cite{Anelli*20}. 
Finally, for our solution using GCNN and its variation using GAT, we applied a linear head with 16 on the former and 100 hidden units on the latter. 
Moreover, for BERT FT, MLP, CNN, and GCNN we used ADAM \cite{kingma2014adam} as optimizer with initial learning rate $1 e^{-4}$, $1 e^{-2}$, $1 e^{-1}$ and $1 e^{-1}$ respectively.
For the GAT experiment, we used SGD as the optimizer with an initial learning rate $3e^{-1}$ and momentum $9 e^{-1}$.
For all the experiments we considered $256$ as batch size, a learning rate scheduler that reduced the learning rate on a plateau with a scaling factor of $1 e^{-1}$, and early stopping.

\begin{table}[h]
\caption{Log-log distribution of the number of posts against the common centralities}
  \caption{Results obtained applying our network-based deep learning models and comparison with state-of-the-art methods }
  
  \label{tab:results}
  \resizebox{1\linewidth}{!}{\begin{tabular}{lccccccc}
    \toprule
    Architecture &Acc&Prec&Recall&AUC$_{ROC}$&AUC$_{PR}$&F1\\
    \midrule
    BERT FT &0.50&0.51&0.50&0.49&0.64&0.50\\
    MLP &0.67&0.67&0.68&0.74&0.74&0.67\\
    CNN &0.70&0.70&0.70&0.77&0.76&0.70\\
    XGBoost\cite{Anelli*20} &0.72&0.72&0.72&0.80&0.80&0.72\\
    \midrule
    TweetGage(GAT) &0.87&0.87&0.87&0.92&0.90&0.88\\
    TweetGage &\textbf{0.89}&\textbf{0.89}&\textbf{0.89}&\textbf{0.95}&\textbf{0.94}&\textbf{0.89}\\
  \bottomrule
  \end{tabular}}
\end{table}

As expected, the fine-tuning of the BERT model on exclusively the corpus of the tweet had the worst performance between the baseline architecture. Therefore, we stated that the features obtained from the text alone are not enough to complete the task we were addressing. 
Looking at the other three baselines instead, we noticed that they were characterized by similar performance with the best results achieved by the model winner of the challenge \cite{Anelli*20}. This result can be explained by considering the fact that the combination of only the features related to the post and the vectorial representation of the text is not enough to predict the impact of a post.
Focusing instead on our proposal and its variation based on the attention mechanism, we saw a significant improvement between $15\%$ and $20\%$ over all the metrics.
The obtained results show that the features derived from the relations between posts in the graph are fundamental in predicting the success of a tweet.

\subsection{Ablation Study}
\label{sub:Ablation}

As a further study, we decided to ablate the importance of the features used in our method. In the previous sections, we presented two different typologies of features for a post $t_i$: those directly related to it, like the number of hashtags or its length (i.e., $\Phi_{t_i}$), and the embeddings of the text contained in the post extracted using a pre-trained BERT model (i.e., $\omega(text_{t_i}$)). 
In our approach, we combined these features with the ones related to the structure of the tweet graph as captured by the Graph Neural Network.

To study the role of the features above, as a preliminary experiment, we excluded the features of $\Phi_{t_i}$ and $\omega(text_{t_i})$. We validated the importance of combining post-related features with the graph neural model. Indeed, from the first row of Table \ref{tab:ablation}, it is possible to observe that only using the information obtained from the connection between the posts, i.e., the features derived from the graph, it is not possible to obtain accurate predictions of the engagement of a post. In fact, the obtained model performed poorly according to all the considered metrics.

In light of this first result, we alternatively considered the features from $\Phi_{t_i}$ and $\omega(text_{t_i})$. The inclusion of these features significantly increased the performance of the obtained model over all the metrics. 

As a final experiment, due to the large dimension of the embeddings obtained using the pre-trained BERT, we applied a dimensionality reduction strategy using the Principal Component Analysis (PCA). Specifically, we used $N=48$ projected features covering more than the $80\%$ of the variance of the features obtained from $\omega(text_{t_i})$. However, this strategy did not increase the model's performances, which maintained an $84\%$ of accuracy.  

\begin{table}[h]
  \caption{Ablation study on the importance of the features considered for our method. $^*$ Feature set reduced with PCA}
  \vspace{+0.4cm}
  \label{tab:ablation}
  \resizebox{1\linewidth}{!}
      {\begin{tabular}{ccccccccc}
        \toprule $\Phi_{t_i}$&$\omega(text_{t_i})$&Acc&Prec&Recall&AUC$_{ROC}$&AUC$_{PR}$&F1\\
        \midrule
         \xmark &\xmark & 0.50 & 0.50 & 0.25 & 0.75 & 0.50 & 0.33\\
         \cmark &\xmark & 0.87 & 0.87 & 0.88 & 0.93 & 0.92 & 0.87\\
         \xmark &\cmark &0.88&0.88&0.88&0.94&0.92&0.88\\
         \xmark &\cmark$^*$ &0.84&0.85&0.85&0.92&0.89&0.85\\
         \cmark$^*$ &\cmark$^*$ &0.87&0.87&0.88&0.94&0.91&0.87\\
         \cmark &\cmark$^*$ &0.88&0.88&0.89&0.95&0.94&0.88\\
         \cmark &\cmark &\textbf{0.89}&\textbf{0.89}&\textbf{0.89}&\textbf{0.95}&\textbf{0.94}&\textbf{0.89}\\ 
      \bottomrule
    \end{tabular}}
\end{table}

Finally, we combined both the $\Phi_{t_i}$ and $\omega(text_{t_i})$ features in three different way: {\em(i)} we reduced the entire feature set using the PCA with $N=48$; {\em(ii)} we reduced the $\omega(text_{t_i})$ with PCA ($N=48$) and kept the $\Phi_{t_i}$ without any projection; and {\em(iii)} we used the entire feature set without any projection. As it is possible to observe, the complete model using the entire feature set obtained the best results in terms of all the considered metrics. This result confirmed the effectiveness of our model by combining both the features extracted by $\Phi_{t_i}$ and $\omega(text_{t_i})$, with a GCNN architecture.

\section{Conclusion}
\label{sec:Conclusion}
In this paper, we proposed a novel approach for the binary classification of posts based on user engagement on Twitter, using Graph Neural Networks.
As a first contribution, our solution introduces a suitable graph-based representation of the relationships between posts published on a social network.
In particular, in our design, we focused on the content of the posts and used the information contained in their hashtags to build an interaction network among posts.
Our intuition was that the engagement dynamics depend on both the properties of a post (such as its length, the author, its content, and so forth) and also on the connection with other previously published content.
To capture these additional features, we leveraged the Graph Neural Network technology and designed a combined solution exploiting both features directly related to posts and the structure of the interaction network of posts to predict the engagement generated on Twitter.
Through a thorough experimental campaign, we proved the effectiveness of our solution and the improvements introduced with respect to previous related approaches.
However, the results reported in this paper must not be considered as an ending point. Indeed, in the future, we plan to extend our research work in different directions. For instance, we plan to add more engagement classes and pass from a binary to a multi-class classification. In this way, we could predict the level of engagement of a post, and use this information to feed a decision support system capable of identifying the best possible author for writing a post about a specific topic. Moreover, we would like to create a new graph model that considers the users as nodes and different edge-types, encoding information such as the usage of the same hashtags, and then predict the user engagement in those interactions. This new representation could help to create a recommendation system that predicts the top users with whom to carry out discussions on a specific topic for generating engagement.  Of course, in such a context, an analysis of the time variance of modeled relationships could be explored to improve the obtained performance.


\bibliographystyle{plain}
\bibliography{bibliography}

\end{document}